\documentclass[11pt,twocolumn,twoside]{article}
\usepackage{fully3d}

\usepackage{amsmath}
\usepackage{siunitx}
\usepackage{graphicx}
\usepackage{tikz}
\usetikzlibrary{shapes.geometric, arrows, decorations.pathmorphing, hobby}
\usepackage{float}
\usepackage{subcaption}
\usepackage{multirow}

\def\micron{\unit{\micro m}}

\def\indent{
    \hspace{4ex}
    }

\DeclareMathOperator*{\argmin}{arg\,min}

\begin{document}

\title{Optimization-based phase retrieval for material decomposition with multi-energy computed tomography} 

\author[1]{Giavanna~Jadick}
\author[1]{Patrick~La~Rivi\`ere}

\affil[1]{Department of Radiology, University of Chicago, Chicago, IL, USA}

\maketitle
\thispagestyle{fancy}



\begin{customabstract}

\smallskip
Multi-energy CT has long demonstrated its ability to enhance image quality with material decomposition. Yet, it has largely been limited to applications that already have high contrast. More recently, x-ray phase-contrast (XPC) imaging has gained interest for its potential to improve detectability in tasks lacking such contrast. 
Previous work has demonstrated the benefit of combining multi-energy imaging with XPC for material decomposition. While the existing method is promising, its analytical approach requires several approximations that limit its broad applicability, and it is based on projection imaging, requiring separate tomographic reconstruction for three-dimensional (3D) volumes.
We propose a natively 3D optimization-based solution that leverages modern computational advances, namely automatic differentiability, to efficiently solve the multi-energy CT phase retrieval problem with a more accurate forward model.
In a simulation study, we quantitatively demonstrate the improvement of our proposed technique relative to the existing analytical standard, and we affirm the benefits of phase contrast over traditional absorption-only x-ray imaging.

\smallskip

\textbf{\textit{This work paves the way for broader application of multi-energy XPC CT in biomedical research, improving detection of weakly absorptive 3D structures that are challenging to observe with traditional x-ray imaging approaches. }}

\smallskip


\end{customabstract}


\section{Introduction}

Multi-energy computed tomography (CT) has long demonstrated its ability to enhance medical image quality by enabling material decomposition: a post-processing step that leverages the unique energy-dependence of different materials' attenuation coefficients to quantitatively solve for a new basis of material images, often bone and tissue. Yet, like most x-ray-based imaging, it has largely been limited to applications that already have high absorption contrast. More recently, x-ray \textit{phase-contrast} (XPC) CT has gained interest for its potential to improve detectability in tasks that lack such contrast. 
XPC imaging leverages the wave-like behavior of x-rays apparent at high spatial resolution; not only are photons attenuated, they subtly refract, producing interference fringes at the edges of even weakly absorptive media.
In traditional contact imaging, these effects are too small to observe. One simple method to render them detectable is \textit{propagation-based} (PB) imaging. With a sufficiently coherent x-ray source, the detector may be moved a significant distance away from the object, increasing the amplitude and number of interference fringes until they are resolvable. This is possible with both radiographic and tomographic geometries and has been successfully implemented using synchrotron and benchtop microfocus x-ray sources. 

\indent These phase-contrast images are used for \textit{phase retrieval}: recovery of the phase-shifting coefficient ($\delta$) of each voxel, much like CT reconstruction recovers the attenuation coefficient ($\mu$) of each voxel. The phrase more broadly refers to a post-processing step that accounts for the phase-contrast physics mechanism. 
With it, the full volume of a weakly-attenuating structure can be reconstructed, further improving its detectability beyond raw edge enhancement. A key challenge in solving this XPC CT inverse problem is that there are two unknowns---both $\delta$ and $\mu$---which requires multiple, independent measurements for a quantitative solution. 

\indent As these coefficients are energy dependent, the process is reminiscent of traditional multi-energy CT material decomposition. Indeed, previous work has presented a successful multi-energy PB-XPC material decomposition technique \cite{schaff2020material}. 
An essential difference relative to traditional multi-energy CT is that, even with two monochromatic measurements and two basis materials, the full PB-XPC forward model is not analytically invertible, requiring simplifying approximations at the cost of image quality and quantitative accuracy.
The existing method has proven successful relative to pure absorption-contrast material decomposition, but residual image artifacts indicate room for improvement, especially in imaging scenarios where the assumptions of its approximations are not satisfied. Furthermore, the method was developed for the two-dimensional (2D) projection image domain and requires a separate step for tomographic reconstruction when applied to three-dimensional (3D) CT imaging, which may introduce additional artifacts \cite{schaff2022spectral}.

\indent We present a natively 3D optimization-based solution to the multi-energy PB-XPC inverse problem. By leveraging modern computational capabilities such as automatic differentiation, we bypass traditional limitations of the existing analytical solution: namely, the many simplifying assumptions. This novel approach has demonstrated great success for single-energy phase retrieval but, to our knowledge, has yet to be applied to the multi-energy case \cite{du2020three}. Moreover, our method does not need separate tomographic reconstruction, as the optimization is performed directly on the basis material volumes. This work paves the way for broader application of multi-energy XPC CT in biomedical research, improving detection of weakly absorptive 3D structures that are challenging to observe with traditional x-ray imaging.

\section{Methods}

\begin{figure}[t]
    \centering
    \includegraphics[width=0.485\textwidth]{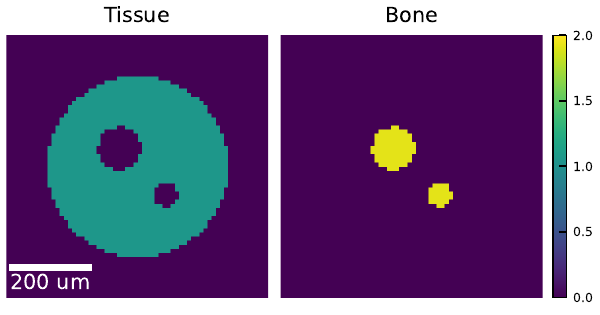}
    \caption{A central slice of the phantom imaged in the simulation experiments. Pixel intensity corresponds to density [g/cm$^3$] of the respective basis material in that voxel. The true density of the soft tissue is 1.06 g/cm$^3$ and of the bone is 1.92 g/cm$^3$.}
    \label{fig:phantom}
  \end{figure}

\subsection{PB-XPC forward modeling}

Any object can be 
defined in terms of its energy-dependent, complex x-ray refractive index $n_E(\mathbf{r}) = 1 - \delta_E(\mathbf{r}) - i\beta_E(\mathbf{r})$, where $\mathbf{r}$ indicates the 3D spatial coordinate $(x,y,z)$ and $E$ is energy. The imaginary part $\mathbf{\beta}$ affects absorption and is related to the linear attenuation coefficient ($\mu = 2k\beta$, where $k$ is the wavenumber), and the real part $\mathbf{\delta}$ produces phase contrast. Then, we define a coherent, monochromatic x-ray plane wave propagating along the $z$-axis as $\psi^{\text{(in)}}_E(\mathbf{r}_\perp)$, where $\mathbf{r}_\perp$ is the 2D spatial coordinate $(x,y)$. 

\indent Transmission of this incident wave field through the object can be modeled in a variety of ways \cite{paganin2021x}. In this case, we implemented the projection approximation, which assumes the object is thin enough that refraction within it is negligible. This is a reasonable first approximation for micro-CT but can be improved upon in the future with more accurate models such as the multislice approach \cite{du2020three}. Then, the wave exiting the object $\psi^{\text{(obj)}}_E(\mathbf{r}_\perp)$ is:
\begin{align}
    \psi^{\text{(obj)}}_E(\mathbf{r}_\perp) &=
    \psi^{\text{(in)}}_E(\mathbf{r}_\perp)
        \exp \left[
            - \int \mu_E(\mathbf{r}) dz
            - ik_E \int \delta_E(\mathbf{r}) dz
        \right].
    \label{eq:pa}
\end{align}
For $\delta = 0$, this reduces to the familiar Beer-Lambert law for traditional absorption-contrast x-ray imaging.

\indent To amplify phase effects to be resolvable by the detector, this object exit wave is then propagated through free space over distance $R$. This can be modeled as a convolution with the Fresnel operator $h_{R,E}(\mathbf{r}_\perp)$:
\begin{align}
    h_{R,E}(\mathbf{r}_\perp) &= \exp \left[
        - \frac{i k_E}{2 R} (x^2 + y^2)
    \right].
    \label{eq:fsp}
\end{align}
The measured signal is equal to the intensity of the wave field at the detector:
\begin{align}
    \mathbf{y}_{\theta} &= \left| 
        h_{R,E}(\mathbf{r}_\perp) * \psi^{\text{(obj)}}_E(\mathbf{r}_\perp)
    \right|
    \label{eq:meas}
\end{align}
where $\theta$ indicates the CT view angle. To acquire the full CT dataset $\mathbf{y}$, the coordinate system is rotated some angle $d\theta$ about the $y$-axis such that the propagation axis $z$ is directed along a new view through the object, and the projection process is repeated.
When applying this model to numerical simulations, one should also account for detector nonidealities such as noise and spatial blur. 
This forward model can be extended to the polychromatic case by simply integrating over the intensity-weighted monochromatic components of the incident plane wave \cite{paganin2021x}. For simplicity, we focus here on the monochromatic case.

\subsection{Multi-energy phase retrieval}

An analytical solution to the multi-energy phase retrieval problem for material decomposition has been developed previously \cite{schaff2020material}. In summary, one first assumes an object is composed of $M$ materials, typically two, then simplifies the integral in Eq.~\ref{eq:pa} to a sum over the refractive indices and thickness profiles of each basis material. This is the same as in analytical absorption-only material decomposition \cite{alvarez1976energy}. For PB-XPC, one also replaces Eq.~\ref{eq:fsp} with the transport-of-intensity equation (TIE) linearized with the assumptions of a weak intensity gradient and near-field distance regime. 
This method has demonstrated success, indicating the potential benefits of material decomposition with PB-XPC, but its applications are limited  by its simplifying assumptions. Furthermore, it is not natively 3D and requires a separate step for tomographic reconstruction when implemented in CT imaging \cite{schaff2022spectral}.

\indent To avoid these limitations, we propose an optimization-based technique to directly reconstruct 3D volumes of basis material densities.
The problem was defined as:
\begin{align}
\label{eq:opt}
    \hat{\mathbf{x}} &= \argmin_{\mathbf{x}} 
    \left(
        || \mathbf{y} - f(\mathbf{x}) ||_2^2
        + \sum_{m} \left[ 
            \tau_m | \mathbf{x}_m |_1 + \gamma_m TV(\mathbf{x}_m) 
        \right]
    \right)
    ,
\end{align}
where $f$ is the forward model outlined in the previous section, $\mathbf{x}$ is an array of the 3D density distribution of each basis material, $m$ is the material index, and $TV$ is the anisotropic total variation. 
The first term is a fidelity term, and the latter two are regularization terms with separate weights $\tau_m, \gamma_m$ for each basis material. 
By allowing for unique weights for each material, Eq.~\ref{eq:opt} accounts for their different physical attributes. 
Also note that for forward modeling, $\mathbf{x}$ is readily converted into $\delta_E(\mathbf{r})$ and $\mu_E(\mathbf{r})$ using the known atomic composition of the basis materials.
The solution was subjected to the non-negativity constraint $\mathbf{x} \geq 0$.

\begin{figure}
    \centering
    \includegraphics[width=0.48\textwidth]{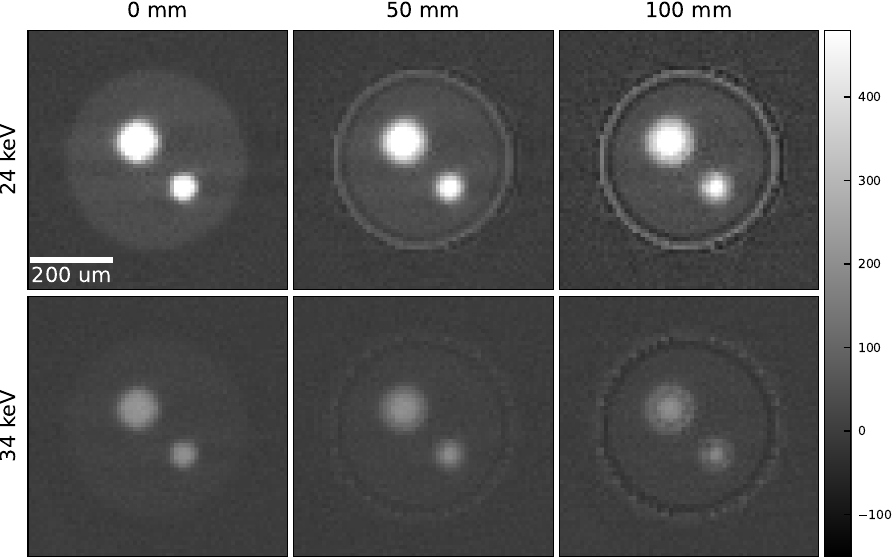}
    \caption{FBP reconstructions of the raw single-energy CT acquisitions (24 and 34 keV) for each propagation distance (0, 50, and 100 mm) prior to phase retrieval. Pixel intensities are in units of effective attenuation (m$^{-1}$).}
    \label{fig:recons}
  \end{figure}

\begin{figure}
    \centering
    \includegraphics[width=0.48\textwidth]{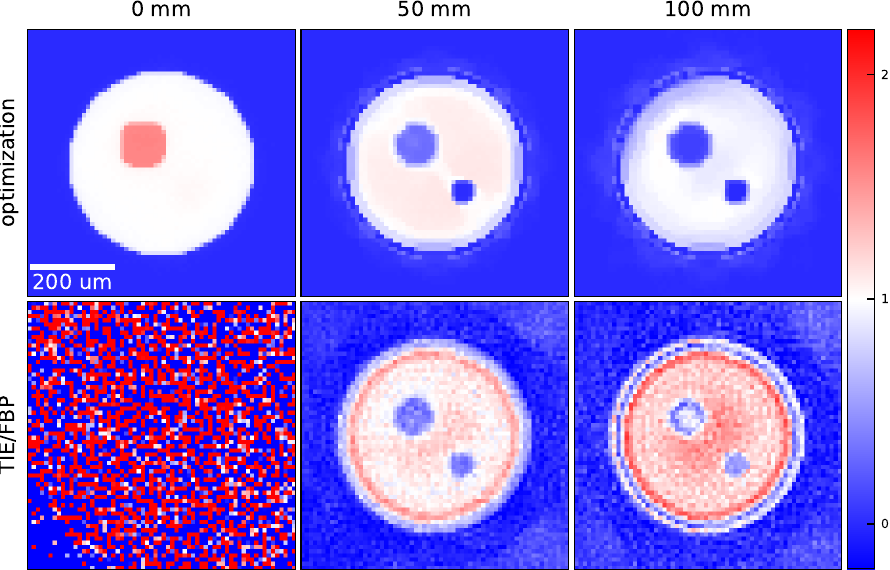}
    \caption{Tissue basis material images reconstructed using optimization (top) or TIE/FBP (bottom) for the three propagation distances. Pixel values are normalized to the true density (1.06 g/cm$^3$), so that 1 indicates a correct result within the true volume.}
    \label{fig:mat1}
  \end{figure}

  \begin{figure}
    \centering
    \includegraphics[width=0.48\textwidth]{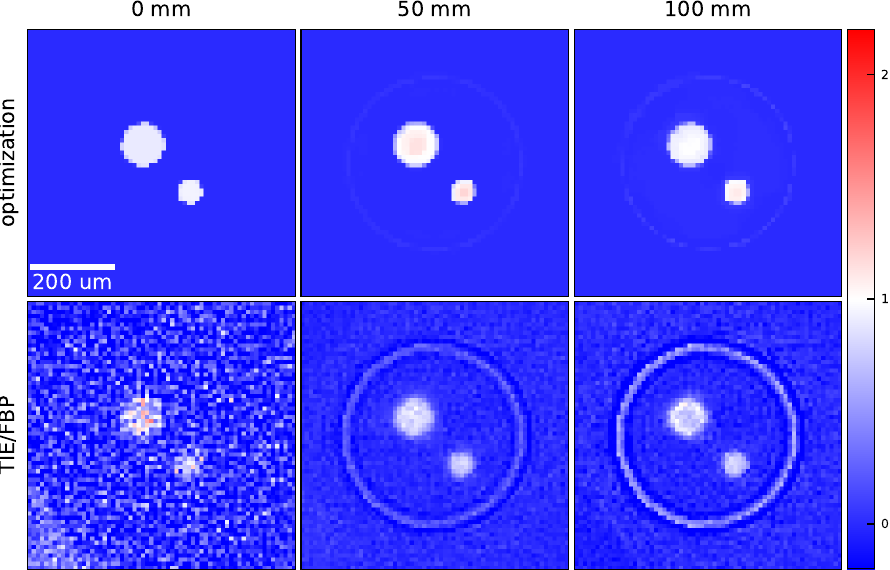}
    \caption{Bone basis material images reconstructed using optimization (top) or TIE/FBP (bottom) for the three propagation distances. Pixel values are normalized to the true density (1.92 g/cm$^3$), so that 1 indicates a correct result within the true volume.}
    \label{fig:mat2}
  \end{figure}

\subsection{Simulated experiment}

Multi-energy PB-XPC CT simulations were performed to assess the performance of the proposed phase retrieval technique. 
Acquisition settings were chosen to emulate parallel-beam synchrotron micro-CT, with monochromatic beam energies of 24 and 34 keV and propagation distances of 0, 50, and 100 mm. A 64$\times$64 detector was simulated with 10-\micron~pixels, a Lorentzian point-spread function with 10-\micron~full-width at half-max, and Poisson counting noise.
A simple two-material phantom was designed to assess the performance: a half-mm cylinder of ICRU soft tissue containing two embedded spheres of ICRU bone (Fig.~\ref{fig:phantom}).

\indent The simulator was written in Python using \verb|chromatix|, a differentiable wave optics library built with \verb|jax| \cite{chromatix_2023}. Its automatic differentiability and native multi-GPU support was leveraged to efficiently solve the optimization problem. A preliminary scan over different optimization parameters led us to choose the Adam optimizer with a learning rate $\alpha = 0.04$, $\tau_1=5 \times 10^{-6},~ \tau_2=10^{-4},~ \gamma_1=10^{-5},$ and $\gamma_2=10^{-5}$. 
For each propagation distance, basis material volumes were reconstructed using both (1) our proposed 3D optimization-based method and (2) filtered back-projection (FBP) of the TIE-based approach \cite{schaff2020material}. As a quantitative comparison, structural similarity index (SSIM) and root-mean-square-error (RMSE) were computed relative to the ground truth in a central slice of each reconstructed material volume.

\section{Results}

Figure \ref{fig:recons} shows FBP reconstructions of the single-energy CT datasets used for the material decomposition. These reconstructions do not attempt to invert any phase-contrast effects, and thus one can see the hallmark edge-enhancement phenomenon of PB-XPC. With propagation distance $R=0$ mm, it is challenging to identify the water cylinder, especially in the 34-keV image. As $R$ increases, the contrast within the cylinder's volume remains similar, but the cylinder itself becomes clearly detectable due to high- and low-intensity fringes at its edges. The bone spheres are easier to identify in all images due to their greater absorption contrast. For both materials, however, there is still blurring at the edges of structures. Phase retrieval aims both to precisely define structure edges and to accurately quantify material density within each structure.

\indent Figures \ref{fig:mat1} and \ref{fig:mat2} show the recovered basis material images for tissue and bone, respectively, using the two phase retrieval methods. 
When $R>0$ mm, the TIE/FBP approach successfully reduces raw image noise, but it has trouble reconstructing internal voids in the tissue image and suffers from edge enhancement artifacts at the bounds of the cylinder, creating a false ring in the bone images. This is consistent with the assumptions of the linearized TIE, which can underestimate PB-XPC edge enhancement effects. The TIE/FBP bone images are more successful, though they have some blurring.
The optimization-based approach successfully recovers the bone image at all $R$, with sharp edges and only a subtle residual ring at the tissue cylinder edge as $R$ increases. 
The tissue images are also successful when $R>0$ mm, with only slight blurring at cylinder edges. 
Without phase contrast ($R=0$ mm), the optimization approach is unable to recover the internal voids.
This indicates the benefit of phase contrast for visualizing weakly attenuating structures; the tissue cylinder is barely visible in the noisy raw images.

\indent To complement the images, Table \ref{tab:metrics} presents the SSIM and RMSE measurements. 
For all distances and materials, the proposed optimization-based technique yielded a higher SSIM and lower RMSE, indicating better qualitative resemblance and quantitative accuracy. The optimization approach's SSIM was maximized with $R=0$ mm, whereas the TIE approach's SSIM was maximized with $R=50$ mm. 
RMSE for the optimization approach was similar at all propagation distances, whereas it was clearly minimized at $R=50$ mm for the TIE approach.
This is consistent with the linearized TIE's assumption that $R$ is in the near-field regime---0 mm is too short, and 100 mm is somewhat too far. The best $R$ for the TIE method is also expected to depend on spatial resolution and energy \cite{paganin2021x}.
Further, for the optimization-based technique, the SSIM measurements invite the question of whether $R>0$ mm is even necessary for good material decomposition. However, Fig.~\ref{fig:mat1} clearly demonstrates that at $R=0$ mm, the technique fails to recover the tissue cylinder's internal voids. This reflects a limitation of the SSIM rather than of phase contrast, and alternative metrics may be useful for future quantification.

\begin{table}
    \centering
    \begin{tabular}{crcccc}
        \toprule
        && \multicolumn{2}{c}{\textbf{SSIM}} &  \multicolumn{2}{c}{\textbf{RMSE}} \\
       \textit{Material} & $R$~~~ & \textit{Opt} & \textit{TIE} & \textit{Opt} & \textit{TIE}  \\
    \midrule
    \midrule
    \multirow{3}{*}{\textbf{Tissue}} 
        & 0 mm &       0.82  &     0.00  &     0.25  &     5.00  \\
        & 50 mm &      0.63  &     0.34  &     0.20  &     0.22  \\
        & 100 mm &     0.54  &     0.22  &     0.21  &     0.36  \\
    \midrule
    \multirow{3}{*}{\textbf{Bone}} 
        & 0 mm &       1.00  &     0.06  &     0.02  &     0.32  \\
        & 50 mm &      0.94  &     0.33  &     0.04  &     0.11  \\
        & 100 mm &     0.83  &     0.21  &     0.04  &     0.17  \\
    \bottomrule
    \end{tabular}
    \caption{Image quality metrics measured in tissue (Fig.~\ref{fig:mat1}) and bone (Fig.~\ref{fig:mat2}) material images for each distance $R$, reconstructed using either the optimization-based or TIE/FBP phase retrieval.}
    \label{tab:metrics}
\end{table}

\section{Conclusions}

We demonstrate the benefit of a natively 3D optimization-based phase retrieval technique for multi-energy PB-XPC CT, quantifying its improvement in basis material image quality and accuracy relative to the current analytical standard. 
As the full PB-XPC forward model is neither analytically invertible nor differentiable, recent computational advances are necessary to enable an efficient optimization-based solution like the one proposed here. 
This was done by writing our simulation framework to utilize the automatic differentiability of \verb|jax|. Furthermore, our technique directly reconstructs 3D material volumes, bypassing the current need for separate tomographic reconstruction.

\indent There are many paths forward for further improvements of our technique. The promise of our simulation should be validated with real experiments and a wider range of parameters, such as smaller detector pixels and larger fields-of-view.
These scenarios are especially challenging for the TIE-based approach and present a unique research need.
In this initial experiment, we considered only a small 64$\times$64 detector to improve simulation and optimization run time. A more realistic detector might have near 1000$\times$1000 pixels. Further, with \verb|chromatix|, we can continue to hone the quantitative accuracy of our forward model while preserving its automatic differentiability \cite{chromatix_2023}.
The optimizer's parameters and regularization weights can also be more rigorously tuned for different imaging tasks and acquisition settings. These improvements will require greater computational expense, and we will investigate ways to accommodate this need using modern computer architectures.
Lastly, a valuable practical improvement will be the inclusion of polychromatic source spectra. When combined with an energy-discriminating detector, such as a spectral photon-counting detector, a polychromatic source can yield multi-energy information in a single shot---the gold standard for multi-energy CT imaging. 

\indent Our ultimate goal is for PB-XPC CT to be a widely available tool for the benefit of biomedical imaging. Many fields could benefit from a convenient 3D imaging technique for visualizing weakly absorptive structures: a challenging task for traditional attenuation-based CT imaging.
Overall, our proposed technique is promising for this purpose and will continue to be refined for future application in experimental work.

\printbibliography

\end{document}